\documentstyle[epsf,prd,twocolumn,aps,floats]{revtex} 
\tightenlines
\begin{document}

\draft
\topmargin = -0.6cm
\topmargin = -2.0cm
\overfullrule 0pt
\twocolumn[\hsize\textwidth\columnwidth\hsize\csname
@twocolumnfalse\endcsname

\title{ 
\vglue -0.5cm
\hfill{\small IFT-P.098/2000} \\
\hfill{\small hep-ph/0103096 }\\
\vglue 0.5cm
See-saw tau lepton mass and 
\\ calculable neutrino masses in a 3-3-1 model}

\author{\bf J. C. Montero$^1$~\footnote{E-mail address: 
montero@ift.unesp.br}, C. A. de S. Pires$^2$~\footnote{
E-mail:cpires@ift.unesp.br}, V. Pleitez~\footnote{E-mail address:
vicente@ift.unesp.br}  }
\address{
$^1$ Instituto de F\'\i sica Te\'orica\\
Universidade Estadual Paulista\\
Rua Pamplona, 145\\ 
01405-900-- S\~ao Paulo, SP\\
Brazil\\
Departamento de F\'\i sica, Universidade Federal da Para\'\i ba\\
Caixa Postal 5008, 58051\\
Jo\~ao Pesoa, Para\'\i ba--PB, Brazil}
\date{\today}
\maketitle
\vspace{.5cm}

\hfuzz=25pt
\begin{abstract}
In this work we show that in a version of the 3-3-1 model proposed by Duong and Ma,
in which the introduction of a scalar sextet is avoided by adding a singlet
heavy charged lepton, the $\tau$ lepton gains mass through a see-saw--like
mechanism. We also show how to generate neutrino masses at the 1-loop level, and
give the respective Maki-Nakagawa-Sakata mixing matrices for a 
set of the parameters. We also consider the effect of adding a singlet
right-handed neutrino. 
\end{abstract}


\vskip2pc]

\narrowtext

Some years ago it was proposed a model with 
$SU(3)_C\otimes SU(3)_L\otimes U(1)_N$ gauge symmetry, 3-3-1 for short, 
in which the three lepton 
families transform in the same way under the gauge symmetry i.e.,
$\Psi_{aL}=(\nu_{a}, l_{a},l^c_a)_L\sim({\bf1},{\bf3},0)$ with $ a=e,\mu,\tau$
\cite{pp,pffoot}.
Therefore, in this model the lepton mass term transforms as 
$({\bf1},{\bf3},0)\otimes({\bf1},{\bf3},0)=
({\bf1},{\bf3}^*,0)_A\oplus({\bf1},{\bf6},0)_S$, hence, in order to give
mass to the charged leptons it is possible to introduce   
a triplet $\eta=(\eta^0,\eta^-_1,\eta^+_2)^T\sim({\bf1},{\bf3},0)$ and a 
symmetric sextet $S\sim({\bf1},{\bf6},0)$~\cite{pffoot}. 
With the $\eta$-triplet only, 
one of the charged leptons remains massless and the other two are mass 
degenerated. Hence, at least a sextet $S$ has to be introduced in order to 
give arbitrary masses to all charged leptons. Although this implies that 
the model has a rather complicated Higgs scalar sector, it is interesting 
to stress that all the extensions of the electroweak standard model 
with extra Higgs scalar multiplets, like
multi-Higgs doublets~\cite{higgshunter}, singlets~\cite{majorons,zee}, 
and triplets~\cite{triplet}, are embedded in this 3-3-1 model. 
In fact, under the subgroup $SU(2)_L\otimes U(1)_Y \subset 
SU(3)_L\otimes U(1)_N$ the model has three scalar doublets,  
four singlets (one neutral, one singly charged and two doubly 
charged), and a complex triplet.
This sort of models also gives some insight on the families replication and
on the observed value of the weak mixing angle $\sin^2\theta_W<1/4$.
Moreover, recently the 3-3-1 models are interesting possibilities for explaining
the new value of the positive muon 
anomalous magnetic moment, reported by the Muon $(g-2)$ 
Collaboration~\cite{g-2}. The new $(g-2)$ value 
is consistent not only with supersymmetry~\cite{susy} but also
with several versions of the 3-3-1 model as it was shown in Ref.~\cite{vietnam}.
Finally, in a version of the model it is possible to generate the top and bottom
quark masses at the tree level while the other quarks and charged leptons
gain mass at the 1-loop level~\cite{mdt}.

The see-saw mechanism was proposed 
in order to understand the smallness of 
the neutrino masses~\cite{seesaw}. On the other hand, later, it was 
introduced a generalization  of this mechanism, which is also valid 
for the charged lepton masses, 
in the context of the minimal left-right symmetric 
model~\cite{davidson,simoes}. Here we will show, in a
3-3-1 model, that a similar mechanism
can be implemented for the case of the $\tau$ lepton mass and that 
the three neutrinos gain mass radiatively. In the case of three neutrinos
the masses and mixing are almost completely determined by the charged lepton
parameters. In fact they depend only upon a coupling constant ($\lambda$) of a
quartic term of the scalar potential. It implies that we are also calculating the
ratio of the neutrino masses (see Eq.~(\ref{bu}) below).

Some years ago, Duong and Ma~\cite{ema} noted that a way to give mass 
to the charged leptons that does not involve a scalar sextet can be 
implemented: 
Besides the usual scalar triplets: 
$\eta$, mentioned before, $\rho=(\rho^+, \rho^0, 
\rho^{++})^T\sim({\bf1},{\bf3},1)$ and
$\chi=(\chi^-,\chi^{--},\chi^0)^T\sim({\bf1},{\bf3},-1)$, 
they introduced a charged lepton transforming like singlet: 
$E^\prime_L\sim({\bf1},{\bf1},-1)$ and  $E^\prime_R\sim({\bf1},{\bf1},-1)$.
In this case the Yukawa interactions are given by
\begin{eqnarray}
-{\cal L}_Y&=&\frac{1}{2}\,\sum_{\stackrel{a,b=e,\mu,\tau}{i,j,k=1,2,3}}
\epsilon^{ijk}\,\overline{(\Psi_{aLi})^c}\,F_{ab}
(\Psi)_{bLj}\eta_k\nonumber \\ &+&
\sum_{a=e,\mu,\tau}[f_a\,\overline{(\Psi)}_{aL}\,E^\prime_R\, 
\rho+f'_a\,\overline{E^\prime_L}\,\chi^T(\Psi^c)_{aR}]
\nonumber \\ &+&M \bar{E^\prime}_LE^\prime_R+H.c.,
\label{yuka}
\end{eqnarray}
where $F_{ab}$ denotes an antisymmetric matrix. Then the mass matrix for the
charged leptons, in the basis of the symmetry eigenstates that we denote 
$l'_{L,R}=(e',\mu',\tau',E')_{L,R}$, reads $\bar{l}'_L{\cal M}^ll'_R+H.c.$ 
where

\begin{equation}
{\cal M}^l=\frac{1}{\sqrt2}\,
\left(\begin{array}{cccc}
0        \,&\,  -f_{e\mu}v_\eta \,  &\, -f_{e\tau}v_\eta & f_ev_\rho \\
f_{e\mu}v_\eta\, & \,      0 \,     &\, -f_{\mu\tau}v_\eta & f_\mu v_\rho\\
f_{e\tau}v_\eta\, & \,  f_{\mu\tau}v_\eta\, & \,    0 & f_\tau v_\rho\\
f'_e v_\chi\, & \, f'_\mu v_\chi\, & \, f'_\tau v_\chi &\sqrt{2}\, M      
\end{array}\right).
\label{mcl}
\end{equation}
Here for simplicity we have assumed real vacuum expectation values:
$\langle\eta^0\rangle\equiv v_\eta/\sqrt2$, $\langle\rho^0\rangle\equiv 
v_\rho/\sqrt2$ and
$\langle\chi^0\rangle\equiv v_\chi^0/\sqrt2$.  

Firstly, by assuming that the only vanishing dimensionless parameters 
in Eq.~(\ref{mcl}) are $f_{e\mu}, f_{e\tau},f_e$ and $f'_e$, we obtain
$m_e=0$ and approximately
\begin{eqnarray}
m_\mu&\approx&-\frac{f_\mu f_{\mu\tau}}{\sqrt2 f_\tau}v_\eta,\quad
m_\tau\approx -\frac{f_\tau f^\prime_\tau v_\rho v_\chi}{2M}-m_\mu,
\nonumber \\ &&\mbox{} m_E \approx M -m_\tau+m_\mu.
\label{taumass}
\end{eqnarray}
To get a positive mass for the $\mu$ and $\tau$ leptons we can redefine the
respective field by a $\gamma_5$ factor i.e., $\mu(\tau)\to\gamma_5\mu(\tau)$
(or see below). We see from Eq.~(\ref{taumass}) that the $\tau$ lepton mass is
of the see-saw type~\cite{davidson,simoes}, however, as showed below,
the $M$ mass is not related to a grand unification scale. 

The mass matrix in Eq.~(\ref{mcl}) is diagonalized by a 
biunitary transformation $U^{l\dagger}_L{\cal M}^lU^l_R=\hat{{\cal M}^l}$, 
relating the symmetry eigenstates (primed fields) with the mass eigenstates 
(unprimed fields), $l'_L=U^l_Ll_L$ and $l^\prime_R=U^l_Rl_R$, where the mass
eigenstates are denoted by $l_{L,R}=(e,\mu,\tau,E)_{L,R}$.

Notice that even for massless neutrinos, which may result in diagonal 
interactions with the $W$ boson, there are flavor changing neutral and charged 
currents in the interactions in Eq.~(\ref{yuka}). 
It means that although the model has many parameters they are not fixed only
by the mass of the charged leptons but by new interactions too.
In this situation,
we can introduce the unitary matrices $U^l_{L,R}$ in Eq.~(\ref{yuka}) and
leaves that phenomenology determine both, the masses of the Higgs scalars and
the matrix elements of those matrices or, we can determine first those matrices
by given the appropriate mass to the charged leptons and use phenomenology
to determine only the mass of the Higgs scalars. Here we will make the first 
part of the second alternative, i.e., we will determine the mixing matrices 
$U^l_{L,R}$ and left for further studies the phenomenology of the model.

Since we will be concerned below with the neutrino mass phenomenology, we find
appropriate to obtain a set of parameters which are consistent with the actual
values of the charged lepton masses. In this vain we use $U^{l\dagger}_L{\cal
M}^l({\cal M}^{l})^\dagger U^l_L=(\hat{{\cal M}}^{l})^2$ 
and $ U^{l\dagger}_R({\cal M}^l)^\dagger{\cal M}^lU^l_R=(\hat{\cal M}^l)^2$ 
in order to find $U^l_{L,R}$.
The numerical analysis of the eigenvalues of  
${\cal M}^l({\cal M}^l)^\dagger$ (or $({\cal M}^l)^\dagger{\cal M}^l$), gives
the following masses for the charged leptons (in GeV): 
\begin{equation}
\begin{array}{c}
m_e=0.000510985,\;m_\mu=0.105658,\\
m_\tau=1.77703,\;m_E=3007.79,
\end{array}
\label{mcl2}
\end{equation}
which are in good agreement with those of Ref.~\cite{pdg}, if the following
parameters are chosen:  
\begin{equation}
\begin{array}{c}
f_{e\mu}=f_{e\tau} = 0.0006785,
\quad f_{\mu\tau}=0.021,\\
 f_e=0.0004263,\;f'_e=0.001,\; 
f_\mu=0.0481815,\; f'_\mu=0.05,\\
f_\tau=0.1259848,\qquad f'_\tau=0.3;
\end{array}
\label{para}
\end{equation} 
and also $v_\chi=1000$ GeV, $M=3000$ GeV, $v_\eta=20$ GeV and 
$v_\rho=245.186(=\sqrt{246^2-20^2})$ GeV. The left- and right-handed mixing
matrices, up to five decimal places, are given by  


\begin{equation}
U^l_L\sim \left(
\begin{array}{cccc}
-0.99517 & -0.09776  & 0.00836  & 0.00002 \\
0.08889 & -0.86222 & 0.49866 & 0.00276 \\
-0.04154 & 0.49701 & 0.86672 & 0.00724 \\
0.00008 & -0.00122 & -0.00766 &  0.99997 \\ 
\end{array}
\right),
\label{maul}
\end{equation}
\begin{equation}
\tilde{U}^l_R\sim \left(
\begin{array}{cccc}
0.99914 & -0.04130 & 0.00433  & 0.00024  \\
0.04133 & 0.99905 & -0.00754 & 0.01175 \\
0.00395 & -0.00852 & -0.99747 & 0.07053 \\
-0.00100 & -0.01116 & 0.07061 &  0.99744 \\ 
\end{array}
\right).
\label{maur}
\end{equation} 

We have verified that in fact, with the matrices given in Eqs.~(\ref{maul}) and 
(\ref{maur}) with $U^{l\dagger}_L {\cal M}^l\tilde{U}^l_R=\hat{{\cal M}^l}$ 
where $\tilde{U}^l_R=U^l_R\phi$  with $\phi={\rm diag}(1,-1,1,-1)$ and 
$\hat{{\cal M}^l}={\rm diag}(m_e,m_\mu,m_\tau,m_E)$. This 
analysis illustrate how the Duong and Ma singlet lepton works~\cite{ema}
to give the correct mass for the charged leptons.  

\begin{figure}[ht]
\centering\leavevmode
\epsfxsize=200pt
\epsfbox{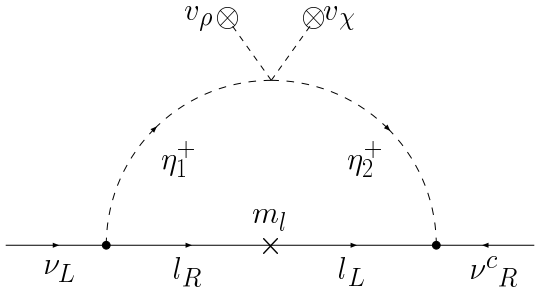}
\vglue -0.01cm
\caption{Diagram showing calculable neutrino masses.}
\label{fig1}
\end{figure}

Until this point neutrinos remain massless. However, there are several
ways to give mass to the neutrinos in the context of the 3-3-1 models. 
These include the mass generation at the tree level~\cite{nradia} or 
by radiative corrections~\cite{radia}. 
Here neutrinos will continue to be massless unless we allow the breakdown of the
$U(1)_{\rm B+L}$ global symmetry in the scalar potential, where $L$ is the total
lepton number and $B$ is the baryon number~\cite{pt1,ma1}. 
This is because if B+L is conserved, the mixture in the singly charged scalar
sector occurs only between $\eta^-_1,\rho^-$, and between $\eta^-_2,\chi^-$, and
diagrams like that in Fig.~\ref{fig1} do not occur, see Eq.(\ref{yuka2}). Recall
that the B+L assignments are 2 for $\eta^-_2$ and $\chi^-$ and 0 for $\eta^-_1$
and $\rho^-$~\cite{ma1}, for leptons B+L coincide with L or with the family
lepton number.
However if B+L is not conserved a quartic term 
$\lambda(\eta^\dagger\chi)(\eta^\dagger\rho) +H.c.$ is allowed and all 
singly charged scalar fields mix one to another.
We will consider the case in which this quartic coupling among the scalar
Higgs does exist and diagrams like that in Fig.~\ref{fig1} do occur. 
The charged current is written in the mass-eigenstate basis as 
$\bar{l}_L\gamma^\mu V_{MNS}\nu_LW^-_\mu$ with the
Maki-Nakagawa-Sakata 
matrix~\cite{mns} defined as $V_{MNS}=(O^l_L)^\dagger U^\nu_L$ where 
$U^\nu_L$ is the matrix which relates the symmetry and the 
left-handed neutrino mass eigenstates. 
Notice that in this case $V_{MNS}$ is a $4\times3$ matrix since $O^l_L$ is 
the submatrix $3\times4$ of $U^l_L$, i.e., omitting
the forth row in Eq.~(\ref{maul}).

Let us now turn on the lepton interactions with the triplet $\eta$:
From Eq.~(\ref{yuka}) we have
\begin{equation}
-\overline{l_{\alpha R}}(O^{lT}_R)_{\alpha a}\,F_{ab}\nu^\prime_{b L}\eta^-_1
-\overline{\nu^{\prime c}_{aR}}\,F_{ab}(O^l_L)_{b\alpha}l_{\alpha L} \eta^+_2.
\label{yuka2}
\end{equation}

Notice that also in Eq.~(\ref{yuka2}) only the $3\times4$ submatrix of 
$U^l_{L,R}$ appear, i.e., omitting the forth row in Eq.~(\ref{maul}) and
(\ref{maur}), since $a,b=e,\mu,\tau$ but $\alpha=e,\mu,\tau,E$; 
we have already denoted those submatrices by $O^l_L,O^l_R$. 

Hence, we have a finite $3\times3$ non-symmetric neutrino mass matrix at the 
1-loop level of the form
\begin{eqnarray}
{\cal M}^\nu_{bd}&\!\!\!=\!\!\!&\lambda v_\rho v_\chi\!\! \sum_{\alpha,a,f}
\!\!(O^{lT}_R)_{\alpha a}F_{ab}m_\alpha F_{df}
(O^l_L)_{f\alpha}\!I(M^2_1,M^2_2,m^2_{\alpha})\nonumber \\ &\!\!\approx\!\!&
\frac{\lambda v_\rho v_\chi}{16\pi^2 M^2_2}\,\sum_{\alpha} m_\alpha\,
(O^{lT}_RF)_{\alpha b}(FO^l_L)_{d\alpha}
\,I(r,s_{\alpha},),
\label{numass1}
\end{eqnarray}
and 
\begin{equation}
I(r,s_{\alpha})=\int^1_0\!\!dx\int^{1-x}_0
\!\!\!\!\!\!\!\!\!\!dy\;\frac{1}{(s^2_{\alpha}-r^2)x+
(s^2_{\alpha}-1)y-s^2_{\alpha}},
\label{inte}
\end{equation}
where $M_{1,2}$ denote typical masses for the singly charged scalars and
we have defined $r=M_1/M_2$ and $s_{\alpha}=m_{\alpha}/M_2$. 
We have neglected a similar contribution proportional to 
$v_\eta^2$, which arises from Fig.~\ref{fig1} if we make $\eta^-_1\to \rho^-$,
$\eta^-_2\to \chi^-$ and $v_\rho v_\chi\to v^2_\eta$, since we are assuming that 
$v^2_\eta\ll v_\rho  v_\chi$. 

\vglue 0.01cm

From Eq.(\ref{numass1}) we can built the symmetric mass matrix $M^\nu=({\cal
M}^\nu+{\cal M}^{\nu T})/2$. 
With $M_1=100$ GeV, $M_2=1$ TeV, and the parameters already obtained for
the charged leptons in Eq.~(\ref{para}), we have from Eqs.~(\ref{numass1}) and
(\ref{inte}) numerical values for $M^\nu$, in GeV and up to a factor $10^{-6}$, 
\begin{equation}
M^\nu\sim \lambda 
\left(\begin{array}{ccc}
-0.00775    & -0.18972 &  0.05030  \\
     & -4.60084 &  1.2635\\
 & & -0.29346
\end{array}\right).
\label{nusmass2}
\end{equation}

This matrix has the following mass eigenvalues: 
\begin{equation}
 \lambda \left(\approx0,\; 0.04931, \;
 4.95892 \right)\,10^{-6}\;{\rm GeV},
\label{bu}
\end{equation}
where the first eigenvalue is assumed to be zero up to the decimal places
we are considering.

This time we diagonalize the mass matrix in Eq.~(\ref{nusmass2}) by making
$\Phi U^{\nu T} M^\nu U^\nu\Phi=\hat{M}^\nu={\rm diag}(m_1,m_2,m_3)$ with
$m_3>m_2>m_1$ where $\Phi={\rm diag}(1,1,i)$ and 
\begin{equation}
U^\nu\sim\left(
\begin{array}{ccc}
-0.99896& -0.02268 & 0.03961 \\
0.03228 & 0.26260 & 0.96436 \\
-0.03228 & 0.96464 & -0.26160  \\
\end{array}\right).
\label{maunusl}
\end{equation}

Notice that the $U^\nu$ matrix does not depend on the global $\lambda$
parameter. In this case we have the mixing matrix $V_{\rm MNS}\equiv
(O^{l}_L)^TU^\nu\Phi$: 
\begin{equation}
V_{\rm MNS}\sim\left(
\begin{array}{ccc}
0.9983 & 0.0058 & 0.0572i \\
0.0538 & 0.2552 & -0.9654i \\
-0.0202 & 0.9668  & 0.2545i \\
-0.0002 & 0.0077 & 0.0008i  \\
\end{array}
\right),
\label{mns}
\end{equation} 
where we have used the basis $l=(e, \mu, \tau, E)^T$ and 
$\nu=(\nu_1,\nu_2,\nu_3)^T$. Although the first row and column of this mixing
matrix is compatible with the SMA mixing matrix, 
we cannot compare directly these numbers with the
usual ones~\cite{gefan,fukugita,kkk} because the matrix in Eq.~(\ref{mns}) is not
unitary, it means that, in the case of three neutrinos (three masses), there are
more parameters than in the case of an unitary mixing matrix (three Euler
angles).  

Now we can try to fix the trilinear $\lambda$ parameter to be consistent 
with the neutrino mass phenomenology: 
From atmospheric neutrino data~\cite{nusat} we know that at least one mass
eigenstate does exist such that its mass is $m>\sqrt{\vert\Delta m^2_{\rm
atm}\vert}\sim (4-6)\cdot10^{-2}$ eV~\cite{smirnov}. On the other hand, the
solar neutrino data requires
$\Delta m^2_{\odot}\sim10^{-5},10^{-7},10^{-10}\,{\rm eV}^2$ depending of the
type of solution LM and SM, LOW or VO solution, respectively~\cite{nussol}. 
Hence by coherence we will choose $\lambda$ in order to be consistent  
with the SMA scenario. 
From Eq.~(\ref{bu}) we see that with $\lambda\sim 6.4\times 10^{-5}$
we obtain $m_1=0$ eV, $m_2=3.2\cdot10^{-3}$ eV and  $m_3=3.2\cdot 10^{-1}$
eV. 

Moreover, we would like to stress that the values of the parameters given in 
Eq.~(\ref{para}) are not expected to be unique and it means that it could be
possible to obtain other values in Eq.~(\ref{nusmass2}) and (\ref{bu}) and of
course, of the MNS-matrix in Eq.~(\ref{mns}), which could fit better the
experimental data~\cite{gefan}. We have verified also that the neutrino masses
and the mixing matrix do not strongly depend on the masses of the charged
scalars $M_1,M_2$. 
In this way, the neutrino parameters are a prediction of the
model: Once the $f$'s constants in Eq.~(\ref{mcl}) are made consistent with 
the known charged lepton masses, the ratios of the neutrino masses
and the mixing matrix $U^\nu$ are also automatically fixed, since $\lambda$ is a
common factor. 
 
Next, we would like to point out that in the 3-3-1 model the
introduction of fermion singlets is more natural than in other models, including
the standard electroweak model, because in the present context the charged
singlet is introduced in order to replace the scalar sextet.
Since the number
of fermion families transforming non-trivially under the 3-3-1 gauge 
symmetry must be a
multiple of 3, the more economic way to introduce new fermions is by adding 
singlets under this symmetry.
Moreover, in the present model which has already a charged lepton singlet, 
if we want to have again a symmetry in the representation content 
between the charged leptons and
neutrinos we can add a right-handed neutrino singlet $N^\prime_R$. In this case 
we can obtain a more wide neutrino mass spectrum if we consider $N^\prime_R$  
as a light neutrino.
In this situation we have the interactions
\begin{equation} 
\sum_{a=e,\mu,\tau}\hat{h}^\prime_a\overline{\psi_{aL}}\,N^\prime_R\eta+
\frac{1}{2}\,
\overline{(N^\prime_R)^c}\,M_R\,N^\prime_R +H.c.
\label{ceci}
\end{equation}

The neutrino mass matrix in the basis $N^\prime=(\nu^\prime_{eL},
\nu^\prime_{\mu L},\nu^\prime_{\tau L}, (N^{\prime c})_L )^T$ and up to
a $10^{-6}$ factor, in GeV, is given by 

\begin{equation}
M^\nu\sim \lambda 
\left(\begin{array}{cccc}
-0.00775    & -0.18972 &  0.05030 & \frac{ \hat{h}_ev_\eta}{2\sqrt2\lambda}  \\
     & -4.60838 &  1.2635 &  \frac{ \hat{h}_\mu v_\eta}{2\sqrt2\lambda}\\
 & & -0.29346 &  \frac{\hat{h}_\tau v_\eta}{2\sqrt2\lambda} \\
  \frac{\hat{h}_e v_\eta}{2\sqrt2\lambda} & 
 \frac{\hat{h}_\mu v_\eta}{2\sqrt2\lambda} &  
  \frac{\hat{h}_\tau v_\eta}{2\sqrt2\lambda} & \frac{M_R}{2\lambda} 
\end{array}\right),
\label{nusmass3}
\end{equation}
where the numerical $3\times3$ submatrix denotes the mass entries generated by
radiative corrections already obtained in Eq.~(\ref{nusmass2}); 
we have defined $\hat{h}_a=10^{6}\hat{h}^\prime_a$. 
The mass matrix in Eq.~(\ref{nusmass3}) has the following mass eigenvalues: 
\begin{equation}
 \left(\approx0,\,0.005,0.052,1.052 \right)\,{\rm eV},
\label{bu2}
\end{equation}
if we chose the following values for the dimensionless parameters
\begin{equation}
(\lambda,\hat{h}_e,\hat{h}_\mu,\hat{h}_\tau)= 
(10^{-6},2\cdot10^{-3},10^{-7},10^{-7}), 
\label{para2}
\end{equation}
and $v_\eta=20$ GeV, $M_R=2$ eV. We can identify, for instance,
$\Delta m^2_{43}=1.1\,{\rm eV}^2$,  $\Delta m^2_{32}=2.7\times10^{-3}\,{\rm
eV}^2$ and $\Delta m^2_{21}=2.1\times10^{-5}\,{\rm eV}^2$ with the respective
mass square difference consistent with
LSND~\cite{lsnd}, atmospheric~\cite{nusat}, and solar~\cite{nussol} neutrino
data. In this case the mixing matrix in the neutrino sector is given by
\begin{equation}
U^\nu\sim \left(
\begin{array}{cccc}
 0.99912 & -0.04199  & -0.00191  & -0.00067\\
-0.04193 & -0.99886& 0.02275 & -0.00091 \\
-0.00293  & -0.02191 & -0.97602 & -0.21654 \\
0 & 0.00582 & 0.21646 & -0.97627 \\ 
\end{array}
\right).
\label{mmisnus}
\end{equation}
The relation between phenomenological neutrinos $N^\prime_a$ and the massive
ones  $N=(\nu_1,\nu_2,\nu_3,\nu_4)^T$ is given by $N^\prime=U^\nu \Phi N$
where $\Phi={\rm diag}(1,i,i,1)$. The $V_{MNS}=U^{l\dagger}_LU^\nu\Phi$ 
mixing matrix is given now by
\begin{equation}
V_{\rm MNS}\sim\left(
\begin{array}{cccc}
-0.9979 & -0.0461i & 0.0445i& 0.0095 \\
-0.0630 & 0.8544i & -0.5048i& -0.1056  \\
-0.0151 & -0.5175i  & -0.8363i & -0.1807\\
-0.0001 & 0.0029i  & 0.2094i  & -0.9778\\
\end{array}
\right).
\label{mns2}
\end{equation} 

Notice that the $3\times3$ (non-hermitian) submatrix involving the known 
charged leptons and the three active neutrinos is again of the SMA solution type
of the solar neutrino problem~\cite{gefan,kkk}. However see the discussion
above. 
In this case we also obtain the mass square
differences which are appropriate for the solar and atmospheric neutrino data.
From the fourth raw and column we see that the heavy charged lepton couples
mainly with the fourth neutrino and the usual charged leptons couple weakly to
this neutrino.
Notwithstanding, also in this case the mixing matrix is not orthogonal and it is
not possible to make a direct comparison with the usual $3\times3$ 
fitting of the data and, as we said before, using a different
set of parameters in Eqs.~(\ref{para}) and (\ref{para2}) we should obtain 
different mixing matrices. 

Then, in this context we have  four phenomenological neutrinos denoted by 
$\nu_{eL},\nu_{\mu L}, \nu_{\tau L},(N^{\prime c})_L$ in the symmetry basis. 
So, in general, we have the possibility of implementing 
``2+2''~\cite{vb1}, ``3+1''~\cite{vb2,kayser,os}, or intermediate neutrino 
mixing schemes.
However it is worth to recall that it is not clear yet that we need four
neutrinos since other type of solutions like flavor changing neutrino
interactions are still possible for the solar neutrino~\cite{guzzo,rzf} and
also for the case of atmospheric neutrino, 
more exotic solutions like neutrino decay and decoherence are not 
ruled out yet~\cite{yasuda}, and moreover, several of these mechanisms can be
working at the same time. 
Notice also that without the sextet we have no chance to assume  
that the charged lepton mass matrix is diagonal, and that the fourth row gives
the couplings of the neutrinos with the heavy charged lepton which are very
small.

Finally, we would like to remember that in the present model 
there are energy independent flavor conversions~\cite{fcnit}. 

\acknowledgments 
This work was supported by Funda\c{c}\~ao de Amparo \`a Pesquisa
do Estado de S\~ao Paulo (FAPESP), Conselho Nacional de 
Ci\^encia e Tecnologia (CNPq) and by Programa de Apoio a
N\'ucleos de Excel\^encia (PRONEX).

\end{document}